\documentclass[twocolumn,aps]{revtex4}

\begin{document}

\title{Classical mechanics in reparametrization-invariant
formulation and the Schr\"odinger equation}
\author{A. A. Deriglazov} \email{alexei.deriglazov@ufjf.edu.br}

\author{B. F. Rizzuti } \email{brunorizzuti@ufam.edu.br}
\altaffiliation{On leave of absence from Instituto de Sa\'ude e
Biotecnologia, ISB, Universidade Federal do Amazonas, AM, Brazil.}

\affiliation{Depto. de Matem\'atica, ICE, Universidade Federal de
Juiz de Fora, MG, Brazil}

\affiliation{Depto. de F\'isica, ICE, Universidade Federal de Juiz
de Fora, MG, Brazil}

\begin{abstract}
The dynamics of every classical-mechanics system can be formulated in
the reparametrization-invariant (RI) form (that is we use the
parametric representation for trajectories, ${\bf x}={\bf
x}(\tau)$, $t=t(\tau)$ instead of ${\bf x}={\bf x}(t)$). In this
pedagogical note we discuss what the quantization rules look like
for the RI formulation of mechanics. We point out that in this
case some of the rules acquire an intuitively clearer form. Hence
the formulation could be an alternative starting point for
teaching the basic principles of quantum mechanics. The advantages
can be resumed as follows. a) In RI formulation both the temporal
and the spatial coordinates are subject to quantization.
b) The canonical Hamiltonian of RI formulation is proportional to
the quantity $\tilde H=p_t+H$, where $H$ is the Hamiltonian of the
initial formulation. Due to the reparametrization invariance, the
quantity $\tilde H$ vanishes for any solution, $\tilde H=0$. So
the corresponding quantum-mechanical operator annihilates the wave
function,  $\hat{\tilde H}\Psi=0$, which is precisely the
Schr\"odinger equation, $i\hbar\partial_t\Psi=\hat H\Psi$. As an
illustration, we discuss quantum mechanics of the relativistic
particle.
\end{abstract}

\maketitle

\section{Introduction}
{\bf Canonical quantization.} The quantum mechanics of a particle can be constructed in several
ways. For instance, Feynman discussed the Schr\"odinger equation
from the Dirac integral representation for $\Psi$. \cite{dd}
Another possible construction is based on physical arguments: the
diffraction of electrons, yielding an interference pattern similar
to that produced by light suggests the emergence of a wave
function governed by a wave equation, which is exactly the
Schr\"odinger equation. \cite{pmrhb, Schiff} We can also
apply the canonical quantization procedure \cite{Schiff, gri}
to a classical mechanical system with the action
\begin{eqnarray}\label{1}
S=\int dt\left[\frac{m}{2}\left(\frac{d{\bf x}}{dt}\right)^2
-V({\bf x}, t)\right],
\end{eqnarray}
where ${\bf x}=(x^1, x^2, x^3)$. To achieve this, we rewrite the
system in Hamiltonian formalism, in terms of the phase-space
variables $x^i, p^i$ equipped with the Poisson bracket $\{x^i,
p^j\}=\delta^{ij}$. The basic quantity now is the Hamiltonian
$H({\bf x}, {\bf p}, t)=\frac{1}{2m}{\bf p}^2+V$. According to the
canonical quantization paradigm, we associate with the phase-space
variables the operators with commutators resembling the Poisson
brackets, $[\hat x^i, \hat p^j]=i\hbar\delta^{ij}$,
\begin{eqnarray}\label{3}
x^i\rightarrow\hat x^i=x^i, \qquad p^i\rightarrow\hat
p^i=-i\hbar\partial_i,
\end{eqnarray}
and postulate on this
base the Schr\"odinger equation for the wave function $\Psi(t, {\bf x})$
\begin{eqnarray}\label{4}
i\hbar\frac{\partial}{\partial t}\Psi=\hat H\Psi, \qquad \hat
H=H(\hat{\bf x}, \hat{\bf p}, t).
\end{eqnarray}
\par
\noindent
{\bf Reparametrization-invariant formulation.}
Anyone classical system can be reformulated in the
reparametrization-invariant (RI) form. To achieve this, we
introduce parametric representation for the trajectory ${\bf
x}={\bf x}(t)$, say ${\bf x}={\bf x}(\tau)$, $t=t(\tau)$, where
$\tau$ is an arbitrary parameter along the trajectory. Denoting
$\frac{d a}{d\tau}\equiv\dot a$, we can write the equalities
$dt=\dot td\tau$, $\frac{d{\bf x}}{dt}=\frac{{}\dot{\bf x}{}}{\dot
t}$. Using these in Eq. (\ref{1}) we obtain the action which is
invariant under the reparameterizations
$\tau\rightarrow\tau'=f(\tau)$
\begin{eqnarray}\label{5}
\tilde S=\int d\tau \tilde L; \qquad \tilde L =\frac{m\dot{\bf
x}^2}{2\dot t}-\dot tV({\bf x}, t).
\end{eqnarray}
This is equivalent to (\ref{1}), as the Lagrangian equations for
the functions ${\bf x}(\tau)$, $t(\tau)$, which follow from
(\ref{5}), imply the correct equations for ${\bf x}(t)$.\cite{aa}
\par \noindent {\it Comment.} We stress that by construction, the
parameter $\tau$, as well as the functions ${\bf x}(\tau)$,
$t(\tau)$ have no direct physical meaning. \cite{green} To
illustrate this, let us consider the free particle, $V=0$. Then
(\ref{5}) implies the equations of motion $(\frac{\dot{\bf
x}}{\dot t}){}^.=0$, $(\frac{\dot{\bf x}^2}{\dot t^2}){}^.=0$. Due
to the reparametrization invariance,  general solution to these
equations contains, besides the integration constants ${\bf v}$
and ${\bf x}_0$, an arbitrary function $g(\tau)$
\begin{eqnarray}\label{5.1}
{\bf x}={\bf v}g(\tau)+{\bf x}_0, \quad t=g(\tau).
\end{eqnarray}
These expressions, while determine the straight line (both in
${\bf x}$ and in $(t, {\bf x})$ spaces), do not specify any
definite evolution law along the line. Only the functions ${\bf
x}(t)$ have the physical meaning. Excluding $\tau$ from the
parametric equations (\ref{5.1}), we obtain ${\bf x}(t)={\bf
v}t+{\bf x}_0$. \par
\noindent
{\bf Formulation of the problem.}
It is the aim of this note to reformulate the quantization rules
listed above for the RI formulation. Our motivations for this type
of presentation of quantum mechanics are as follows. \par
\noindent 1. From Eq. (\ref{3}) there may arise an impression that
in quantum mechanics only the spatial coordinates are subject to
quantization, see, for example, the text-books in Ref.
\onlinecite{sakurai} and \onlinecite{sred}. Moreover, in Ref.
\onlinecite{sred} it is claimed that it is "surprisingly
complicated" to promote time as an operator. As will be shown
below, quantization of the temporal coordinate does not represent
any special problem in the RI formulation. So, the fact that time
is not quantized can be regarded as an artefact of the formulation
used, and does not represent an intrinsic property of the
quantum-mechanics quantization paradigm. \par \noindent 2. In RI
formulation, the quantity that vanishes for any true trajectory of
the phase space naturally appears. The quantum counterpart of this
quantity is precisely the Schr\"odinger equation. That is, the RI
formulation implies (one more) simple intuitive argument for
postulating the Schr\"odinger equation.
\par \noindent 3. Relativistic systems are usually formulated in
the RI form, \cite{gt, mhct, aad1, bm, aadg} so the familiarity with RI
formulation of classical mechanics can be important for proper
understanding of the special-relativity theory.

\section{Quantum mechanics of the RI formulation}
The action (\ref{5}) is defined on configuration space with the
coordinates $x^i, t$. So, to reach the Hamiltonian formulation, we
introduce the phase space parameterized by $x^i$, $p^i$, $t$,
$p_t$, with the Poisson brackets defined as
\begin{eqnarray}\label{6}
\{x^i, p^j\}=\delta^{ij},
\end{eqnarray}
\begin{eqnarray}\label{6.1}
\{t, p_t\}=1.
\end{eqnarray}
According to the standard Hamiltonization prescription
\cite{aad1, gold}, the variables $x^i(\tau)$, $t(\tau)$ obey the
Euler-Lagrange equations, while the dynamics of the conjugate
momenta is specified by the equations
\begin{eqnarray}\label{7}
p^i=\frac{\partial \tilde L}{\partial\dot x^i}=\frac{m\dot
x^i}{\dot t}, \, p_t=\frac{\partial \tilde L}{\partial\dot t}
=-\frac{m\dot{\bf x}^2}{2\dot t^2}-V.
\end{eqnarray}
These equations imply the constraint \cite{foot2}
\begin{eqnarray}\label{8}
\tilde H\equiv p_t+\frac{1}{2m}{\bf p}^2+V=0,
\end{eqnarray}
which is satisfied for any solution to equations
of motion. It reappears once again when we try to construct the
canonical Hamiltonian of the action (\ref{5})
\begin{eqnarray}\label{9}
p^i\dot x^i+p_t\dot t-L=\dot t\left[p_t+\frac{1}{2m}{\bf
p}^2+V\right],
\end{eqnarray}
which thus vanishes for any true trajectory. \par
\noindent
{\it Comment.} Being
reparametrization invariant, the action (\ref{5}) represents an
example of a theory with local symmetry. The appearance of
constraints in the Hamiltonian formalism is a characteristic
property of such theories. A systematic method for analysis of a
locally-invariant theory has been suggested by Dirac, \cite{dirac}
and is now based on solid mathematical grounds, see, for example,
the text-books in Ref. \onlinecite{gt, mhct, aad1}. For the exact
relationship between local symmetries and Hamiltonian constraints
see Ref. \onlinecite{aad1,aadkee, aadsy}.

To quantize the RI formulation, we replace the phase-space
variables by operators that resemble the brackets (\ref{6}), (\ref{6.1})
\begin{eqnarray}\label{10}
\begin{array}{cccccc}
t\rightarrow\hat t=t, & p_t\rightarrow\hat
p_t=-i\hbar\partial_t,  \\
x^i\rightarrow\hat x^i=x^i, & p^i\rightarrow\hat
p^i=-i\hbar\partial_i.
\end{array}
\end{eqnarray}
Since the phase-space function $\tilde H$ vanishes in classical
theory, we expect that the corresponding quantum-mechanical
operator annihilates the wave function, $\hat{\tilde H}\Psi=0$.
Taking into account Eqs. (\ref{8}), (\ref{10}), the condition
reads
\begin{eqnarray}\label{11}
i\hbar\partial_t\Psi=(-\frac{\hbar^2}{2m}\triangle+V)\Psi.
\end{eqnarray}
That is, we have arrived at the Schr\"odinger equation. \par
\noindent
{\it Comment.} In the standard formulation (\ref{1}) the
commutators $[\hat x^i, \hat p^j]=i \hbar \delta^{ij}$, being
combined with the formula
\begin{equation}\label{21}
(\Delta A)^2(\Delta B)^2 \geq (\frac{1}{2i}[\hat A, \hat B])^2,
\end{equation}
for the standard deviation of the hermitian operators $\hat A$,
$\hat B$ (see the derivation on pages 108-109 of Ref.
\onlinecite{gri}), implies $(\Delta x^i)(\Delta p^j) \geq
\frac{\hbar}{2}\delta^{ij}$. By this way, the position-momentum
uncertainty relation arises on a fundamental level, as a pure
algebraic fact. In contrast, it is well known \cite{gri, mt, ab, pb}
that the energy-time uncertainty relation has very different
origin and interpretation. (See also a particular derivation of
energy-time uncertainty relation on page 112 of Ref.
\onlinecite{gri}).

In the reparametrization invariant formulation (\ref{5}) we have,
in addition to (\ref{6}), the commutator $\{\hat t, \hat p_t\}=i\hbar$.
So, one asks whether the energy-time uncertainty relation
can be derived in the same way,  \cite{ack} as an algebraic consequence of (\ref{6.1}).
We point out, that the appearance of the bracket (\ref{6.1})
on an equal footing with (\ref{6}) does not mean complete symmetrization
of the quantum-mechanics formalism with respect to the time and the position variables.

The asymmetry has various origins, some of them are enumerated
below: a) Scalar product implies integration on the position
variables at a fixed instant. b) The operator $\hat p_t$ is
hermitian only on the subspace of solutions of the Shcr\"odinger
equation, see Eq. (\ref{11}). c) The canonically conjugated
variable for $t$ is $p_t$, not $H$. \par
\noindent
{\bf General case.}
The recipe also works for the general case. Bearing in mind
possible applications for the many-particle systems interacting
with an electromagnetic field, let us consider the action
\begin{eqnarray}\label{12}
S=\int dtL\left(q^A, \frac{dq^A}{dt}, t\right),
\end{eqnarray}
where $q^A=({\bf x}_1, {\bf x}_2, \ldots , {\bf x}_n)$, $A=1, 2,
\ldots , 3n$ stand for generalized coordinates of $3n$-dimensional
configuration space of $n$ particles. Supposing that the action is
non-singular, equations for the momenta $p_A=\frac{\partial L(q,
v, t)}{\partial v^A}$ can be resolved algebraically with respect
to $v^A$, $v^A$ $=$ $v^A(q^B, p^C, t)$. Then the physical
Hamiltonian reads
\begin{equation}\label{14}
H(q^A, p_A, t)=p_A v^A-L(q, v, t).
\end{equation}
The RI action reads
\begin{eqnarray}\label{14}
\tilde S=\int dt\tilde L=\int d\tau \dot tL\left(q^A, \frac{\dot
q^A}{\dot t}, t\right).
\end{eqnarray}
This leads to the following equations for conjugate momenta
\begin{eqnarray}\label{15}
p^i=\left.\frac{\partial L(q, v, t)}{\partial v^A}
\right|_{v^A\rightarrow\frac{\dot q^A}{\dot t}},
\end{eqnarray}
\begin{eqnarray}\label{16}
p_t=\left.\left[L(q, v, t)- \frac{\dot q^A}{\dot t}\frac{\partial
L(q, v, t)}{\partial v^A}\right]\right|_{v^A\rightarrow\frac{\dot
q^A}{\dot t}}.
\end{eqnarray}
The equation (\ref{15}) can be solved as $\frac{\dot q^A}{\dot
t}=v^A(q, p, t)$. Using this expression in (\ref{16}), we obtain
the constraint $p_t+p_Av^A-L(q, v, t)=0$, or, equivalently
\begin{eqnarray}\label{17}
p_t+H=0,
\end{eqnarray}
which is satisfied for any solution to the phase-space equations
of motion. Here $H$ stands for the Hamiltonian of the initial
formulation $L$. Similarly to the example discussed above, the
canonical Hamiltonian of the RI formulation (\ref{14}) is
proportional to the constraint
\begin{eqnarray}\label{18}
p_A\dot q^A+p_t\dot t-\tilde L=\dot t\left[p_t+p_Av^A-L\right].
\end{eqnarray}
To quantize the RI formulation, we replace the phase-space
variables by operators, $p_t\rightarrow\hat
p_t=-i\hbar\partial_t$, $p_A\rightarrow\hat
p_A=-i\hbar\partial_A$, and impose the equation (\ref{17}) as a
constraint on the wave function. This immediately leads to the
Schr\"odinger equation $i\hbar\partial_t\Psi=\hat H\Psi$.

\section{Example: Quantum mechanics of the relativistic particle}
Using the physical coordinates ${\bf x}(t)$, the relativistic
particle action reads
\begin{equation}\label{26}
S=-mc\int dt\sqrt{c^2-\Big(\frac{d{\bf x}}{dt}\Big)^2},
\end{equation}
where $m$ is the mass of the particle and $c$ is the speed of
light. Introducing an arbitrary parametrization ${\bf x}(\tau)$,
$t(\tau)$ of the trajectory, the action acquires the RI form
\begin{equation}\label{27}
\tilde S=-mc\int d\tau\dot t \sqrt{c^2-\frac{\dot{\bf x}^2}{\dot t^2}}.
\end{equation}
If we agree to consider only the parameterizations adjusted with
the "time flow", $\frac{dt}{d\tau}>0$, the action can be written
in  the manifestly relativistic-invariant form
\begin{equation}\label{27.1}
\begin{array}{cccccc}
\tilde S=-mc\int d\tau\frac{\dot t}{|\dot t|}\sqrt{(c\dot t)^2-
\dot{\textbf x}^2} \\
=-mc\int d\tau\sqrt{\eta_{\mu\nu}\dot x^\mu\dot x^\nu},
\end{array}
\end{equation}
where the Minkowsky metric has been chosen as  $\eta_{\mu\nu}=(+,
-, -, -)$.

Passing to the Hamiltonian formulation for (\ref{27}), we
introduce the canonical momenta
\begin{equation}\label{29}
{\bf p}=\frac{\partial \tilde L}{\partial \dot{\bf x}}=\frac{mc\dot
{\bf x}}{\dot t\sqrt{c^2-\frac{\dot{\bf x}^2}{\dot t^2}}},
\end{equation}
\begin{equation}\label{28}
p_t=\frac{\partial \tilde L}{\partial \dot
t}=-mc{\sqrt{c^2-\frac{\dot{\bf x}^2}{\dot t^2}}}.
\end{equation}
The first equation can be used to present $\dot{\bf x}$ through
${\bf p}$ and $\dot t$, $\dot{\bf x}=\dot t \frac{c{\bf
p}}{\sqrt{m^2c^2+{\bf p}^2}}$. Using this result in Eq.
(\ref{28}), we obtain the basic constraint
$p_t=-c\sqrt{m^2c^2+{\bf p}^2})$. As it should be, the canonical
Hamiltonian turns out to be proportional to the constraint,
$\tilde H=\dot t(p_t+ c\sqrt{m^2c^2+{\bf p}^2})$.

Quantizing the model via the RI approach, we arrive at the
Shr\"odinger equation which is just the square-root Klein-Gordon
equation \cite{sakurai}
\begin{equation}\label{30.1}
i\hbar\partial_t \Psi=c\sqrt{m^2c^2-\hbar^2\triangle}\,\,\Psi,
\end{equation}
where $\triangle = \frac{\partial^2}{\partial x^2_1}+
\frac{\partial^2}{\partial x^2_2}+ \frac{\partial^2}{\partial x^2_3}$.
For the latter use, we write it in the equivalent form
\begin{equation}\label{30}
\left[i\frac{\partial}{c\partial
t}-\sqrt{\mu^2-\triangle}\,\right]\Psi=0, \quad
\mu\equiv\frac{mc}{\hbar}.
\end{equation}
It implies the right nonrelativistic limit. Indeed, expanding the
square root into power series with respect to $\frac{1}{c^2}$ and
keeping the leading two terms, we observe that the function
$\chi\equiv\exp{(-i\frac{mc^2}{\hbar}t)}\Psi$ obeys the
nonrelativistic Schr\"odinger equation
$i\hbar\partial_t\chi=-\frac{\hbar^2}{2m}\triangle\chi$.

Dealing with the equation (\ref{30}), we are faced with two
well-known problems. First, it contains the square-root operator.
Second, it has no the manifestly relativistic-covariant form. We
demonstrate now that both problems can be avoided, reformulating
the theory in the {\it equivalent form} in terms of the {\it real} scalar field $\phi(x^\mu)$
instead of the complex wave function $\Psi$.

Consider the manifestly relativistic Klein-Gordon equation for the
{\it real function} $\phi$
\begin{equation}\label{30.2}
\left[\partial_\mu\partial^\mu+(\frac{mc}{\hbar})^2\right]\phi=0.
\end{equation}
The equations (\ref{30}) and (\ref{30.2}) turn out to be
equivalent in the following sense: \par \noindent A) If $\phi$ is
a solution to the Klein-Gordon equation (\ref{30.2}), then
\begin{equation}\label{30.3}
\Psi=\Psi_1+i\Psi_2
=-\sqrt{\mu^2-\triangle}\,\,\phi-i\frac{\partial}{c\partial t}\phi,
\end{equation}
obeys the Schr\"odinger equation (\ref{30}). We point out an
analogy with the electrodynamics: as the vector potential ${\bf
A}$ produces the electric and magnetic fields, ${\bf
E}=-\frac1c\partial_t{\bf A}$, ${\bf B}={\bf\nabla}\times{\bf A}$,
the real field $\phi$ produces the real and imaginary parts of the
wave function according to Eq. (\ref{30.3}). So, we call $\phi$
the wave-function scalar potential. \cite{aa2} \par \noindent B)
If $\Psi$ is a solution to the Schr\"odinger equation (\ref{30}),
then the function
\begin{equation}\label{30.4}
\phi=k({\bf x})-c\int_{0}^{t} d\tau\Psi_2(\tau, {\bf x}).
\end{equation}
obeys the Klein-Gordon equation (\ref{30.2}). It has been denoted
\begin{equation}\label{30.5}
k({\bf x})=-\int\frac{d^3p}{(2\pi)^{\frac{3}{2}}}\frac{e^{i{\bf
px}}\Psi_1({\bf p})}{\sqrt{\mu^2+{\bf p}^2}},
\end{equation}
and $\Psi_1({\bf p})$ is the Fourier-transformation of $\Psi_1(0,
{\bf x})$, $\Psi_1(0, {\bf
x})=\int\frac{d^3p}{(2\pi)^{\frac{3}{2}}}e^{i{\bf px}}\Psi_1({\bf
p})$. The function $k({\bf x})$ represents a formal solution to
the equation $\sqrt{\mu^2-\triangle}\,\,k({\bf x})=-\Psi_1(0, {\bf
x})$, the latter is the real part of the equation (\ref{30}) taken
at the instant $t=0$.

According to Eq. (\ref{30.3}), the probability density can be
presented through the wave-function potential
\begin{equation}\label{30.6}
\Psi^*\Psi=\left(\frac{\partial}{c\partial
t}\phi\right)^2+\left(\sqrt{\mu^2-\triangle}\,\,\phi\right)^2.
\end{equation}
In turn, it allows us to identify the probability density with the
energy density of the field $\phi$. Indeed, the equation of motion
(\ref{30.2}) can be obtained from the action
\begin{equation}\label{30.7}
S=\int d^4x\left[\left(\partial_0\phi\right)^2-
\left(\sqrt{\mu^2-\triangle}\,\,\phi\right)^2\right],
\end{equation}
so the right-hand side of Eq. (\ref{30.6}) is just the energy
density of $\phi$.

In resume, quantum mechanics of the relativistic particle can be
reformulated in the manifestly relativistic-covariant form in
terms of the real wave-function potential (\ref{30.2}).
Interaction with the electromagnetic field can be achieved adding
the term
\begin{eqnarray}\label{30.8}
S_{int}=\int dt\left[eA_0+\frac{e}{c}A_i\frac{dx^i}{dt}\right]
=\int d\tau\frac{e}{c}A_\mu\dot x^\mu.
\end{eqnarray}
Repeating the analysis made above, we arrive at the equation
(\ref{30.1}), where one needs to replace
$\partial_i\rightarrow\partial_i-i\frac{e}{\hbar c}A_i$,
$c\partial_t\rightarrow\partial_0-i\frac{e}{\hbar c}A_0$.
Unfortunately, in this case we are not able to reformulate the
theory in the manifestly relativistic-covariant form.

\section{Concluding remarks}
We have shown how the canonical quantization can be reformulated
in a fair way: the time variable stands on an equal footing with
spatial variables, both being quantized. For that, we work with
the reparameterization invariant action, where time and spatial
variables are functions of an arbitrary parameter along the
trajectory. Reparametrization invariance implies the constraint
(\ref{17}) which holds for any true trajectory. The corresponding
quantum-mechanical operator annihilates the wave function, leading
precisely to the Schr\"odinger equation. As an application for the
RI formulation, we demonstrated that the Klein-Gordon equation for
the {\it real field} has the probabilistic interpretation.


\begin{thebibliography}{5}
\bibitem{foot2} The constraint can be written as $p_t=-H$, where
$H$ is the Hamiltonian of the initial formulation. For this
reason, energy is often called the conjugate momentum for the time
variable, see C. Lanczos, {\it The variational principles of
mechanics} (University of Toronto Press, Toronto, 1949), pp.
132-133. Observe also that $H(q, p, t)$ is generally not a
conserved quantity.
\end{thebibliography}
\end{document}